\documentclass[aps,prb,twocolumn,showpacs,floatfix]{revtex4-1}
\usepackage{graphicx}
\usepackage{psfrag}
\usepackage{color}
\usepackage{amsmath}
\usepackage{bm}
\usepackage[toc,page]{appendix}

\begin{document}  

\title{Zero-temperature criticality in the Gaussian random bond Ising model on a square lattice}

\author{Olga Dimitrova}

\affiliation{Institut f\"{u}r Theoretische Physik, Universit\"{a}t zu K\"{o}ln, Z\"{u}lpicher Str. 77, D-50937 K\"{o}ln, Germany}

\date{\today}

\begin{abstract}
The free energy and the specific heat of the two-dimensional Gaussian random bond Ising model on a square lattice are found with high accuracy using graph expansion method. At low temperatures the specific heat reveals a zero-temperature criticality described by the power law $C\propto T^{1+\alpha}$, with $\alpha= 0.55(8)$. Interpretation of the free energy in terms of independent two-level excitations gives the density of states, that follows a novel power law $\rho(\epsilon)\propto \epsilon^\alpha$ at low energies. An exact high-temperature series for this model up to the term $\beta^{29}$ is found. A proof that the density of one-site spin flip states vanishes at low energy is given.
\end{abstract}


\maketitle

\section{Introduction}

Nernst's law of thermodynamics states that the entropy and the specific heat of physical systems must vanish at zero temperature. Experiments, that study the specific heat at low temperatures, are a simple yet powerful tool to reveal the nature of the ground state and of the low-energy excitations. Examples are the Stefan-Boltzmann law, $C\propto T^3$, revealing the relativistic nature of the photons in equilibrium, the specific heat of a metal: $C\propto T$, which arises from the Fermi surface and the Pauli exclusion principle. In the insulating solids the two common laws: $C\propto T^3$ and $C\propto T^{3/2}$, distinguish between the massless phonons and the massive spin-wave excitations. The numerical experiment of this paper finds a novel power law behaviour of the specific heat at low temperatures in a realistic model of a spin glass.

The disordered two-dimensional systems are ubiquitous in the modern experiment, reflecting the current technological state of growing films. Put simplistically, there are two dimensions to disorder: the strength and the frustration it induces in the system, both varying with the chemical composition. Accordingly, there are several types of order at zero temperature: a homogeneous order, when the frustration is small and an alternative order, the random glass state, frequently observed at low temperatures~\cite{BY,HemmenMorgenstern}. In magnetic spin systems with a phase transition in a low-temperature ordered phase, like the ferromagnet, weak disorder suppresses the phase transition temperature. In strongly disordered systems the phase transition in the spin glass state, that normally takes place at a finite freezing temperature, $T_f$, can be suppressed by varying the chemical composition, as well as by decreasing the film thickness~\cite{FisherHuse,KenningSlaughterCowen,HoinesStubiLoloeeCowenBass}, all the way down to zero temperature.

The random bond Ising model, describing disordered spin systems, is the short-range version of the Edwards-Anderson model~\cite{EdwardsAnderson}. Usually, such models are divided into several classes: with continuous and discontinuous, and symmetric and asymmetric, distributions of the bond interactions. For a continuous symmetric disorder, of which the Gaussian distribution is a typical example, the random bond Ising model develops no order in two dimensions~\cite{MorgensternBinder1,McMillanGaussian,BrayMoore,CheungMcMillanGaussian}, with $T_f$ equal to zero. In this situation two-dimensional thermal fluctuations can produce a zero-temperature criticality phenomenon.

At low temperatures the thermodynamic properties of the spin glass are determined by its low-energy spectrum, described qualitatively by the picture of the droplet excitations~\cite{FH}. In particular, this spectrum determines the lower critical dimension. In this paper, for the two-dimensional Gaussian random bond Ising model on the square lattice, we find the free energy and the specific heat in a wide range of temperatures and show, that the density of states vanishes at low energies, unlike the density of states in the spin glass.

In the homogeneous spin systems, where the discrete symmetry, like $Z_2$ for the Ising system, breaks down spontaneously, the ground state [or the pair of ground states] is unique, and all excitations have a finite energy, the gap. Non-frustrating disorder induces tails of the density of states inside the gap. In two dimensions these excitations are the domain walls, encircling the connected area of flipped spins. The longer the domain wall is, the more excitation energy it brings in. Similarly, the ground state of the two-dimensional Gaussian random bond Ising model is probably also unique~\cite{Newman,Arguin}, whereas the low-energy excitations are droplets~\cite{FH,McMillanDroplet,BrayMooreDroplet}, representing connected clusters of flipped spins with fractal-like boundaries. Since the two dimensions lie below the lower critical dimension of the spin glass phase~\cite{BY}, the energy of the droplets, albeit distributed randomly, scales with the droplet size $L$ as $\epsilon\propto L^\theta$, where the exponent $\theta\approx - 0.295$ has been found in multiple 
studies~\cite{BrayMoore,McMillanGaussian,Amoruso,KatzgraberYoungTheta,HoudayerHartmann}. 
The boundary of the droplet (the domain wall) treads a long fractal path before its total energy, locally positive and negative due to the frustration, will be fine-tuned into some positive value close to zero. From the thermodynamic point of view these droplet excitations represent two-level, flip/non-flip states, taken as non-interacting~\cite{AndersonHalperinVarma}. They determine the free energy and the specific heat at low temperatures. The extended low-energy excitations can be studied by methods of classical statistics, despite that at low temperatures the quantum fluctuations dominate on small scales. In the work~[\onlinecite{CheungMcMillanGaussian}] the density of states for a continuous and symmetric distribution of the bond interactions was found to be finite at zero energy and growing linearly with the excitation energy. Accordingly, the specific heat was found to behave linearly, $C\propto T$, at low temperatures~\cite{CheungMcMillanGaussian,Thomas,JLMM}.

In this paper we study the two-dimensional Gaussian random bond Ising model on the square lattice at low temperatures. The usual method of choice for studying such models is the Monte-Carlo simulations. However, the Monte Carlo method at low temperatures suffers markedly from slowing down~\cite{BY}. The graph expansion method~\cite{sikes} gives observables, properly averaged over disorder, in the thermodynamic limit of an infinite lattice in terms of observables of the same model restricted to small clips of the lattice. Owing to a lack of the intervening phase transition from the paramagnetic phase in the spin-glass state and the finite size of the clips, the high-temperature series, defined and valid in the paramagnetic phase at high temperatures, can be continued all the way down towards zero temperature, provided a long enough series is known. In our approach, the Griffiths singularities~\cite{BY}, ubiquitous in the disordered systems, develop progressively as the size of the clips grows and, eventually, a proper average becomes divergent from a typical average as the temperature freezes, whereas the Monte-Carlo simulations and the new Pfaffian algorithms, that have been recently developed in~[\onlinecite{Thomas}] on larger samples, rely on the typical average. In this paper we find the free energy and the specific heat. The specific heat vanishes at low temperatures, in accordance with the third law of thermodynamics, and is found to follow the power law: $C(T)\propto T^{1+\alpha}$, with $\alpha\approx 0.55$. The $3/2$-law has also been observed in the experiments~\cite{BY,HemmenMorgenstern}, though, its origin might be due to the spin-waves, different to that found in our paper. In terms of independent two-level states, we find, that the density of states follows the power law, $\rho(\epsilon)\propto \epsilon^\alpha$, at a low energy $\epsilon$. 

This paper is organized as follows. In section II the Gaussian random bond Ising model on the square lattice is described and the graph expansion method is explained. In section III the proof of the vanishing density of one-site states is given. In section IV the efficient high-temperature series method for evaluating the thermodynamic observables on small clips from the lattice is described. In section V the results for the average free energy and the specific heat are found. In the approximation of independent two-level excitations the density of states is found at low energies.

\section{Model and graph expansion}

The two-dimensional Gaussian random bond Ising model on the square lattice describes an anisotropic easy-axis magnetic system. It consists of Ising spins, characterized by a binary value $s_{\mathbf{x}}=\pm 1$, assigned to each site $\mathbf{x}$ on the square lattice. The two spins across the bond $\langle \mathbf{x} \mathbf{y} \rangle$, a pair of the nearest-neighbour sites $\mathbf{x}$ and $\mathbf{y}$, are coupled by an exchange interaction. The total energy of this system reads
\begin{equation}\label{RBI}
H=-\sum_{\langle \mathbf{x}\mathbf{y} \rangle} J_{\mathbf{x}\mathbf{y}} s_{\mathbf{x}} s_{\mathbf{y}}.
\end{equation}
The nearest-neighbour interaction $J_{\mathbf{x}\mathbf{y}}$ is a quenched random variable, which changes from one bond to another, and is distributed independently, randomly and symmetrically around zero:
\begin{equation}\label{distri}
P[J_{\mathbf{x}\mathbf{y}}]=\frac{1}{\sqrt{2\pi} J}\exp\left(-\frac{J^2_{\mathbf{x}\mathbf{y}}}{2J^2}\right).
\end{equation}
The average strength of the bond interaction can be normalized as: $J=1$, without loss of generality. The randomness of the bond interactions, Eq.~(\ref{distri}), introduces a certain amount of frustration to the spin order. In particular, finding the ground state for a given disorder represents a difficult, global minimization problem. It is difficult to construct a meaningful theory, starting from this unknown ground state. Instead, the proper glass order parameter is of the Edwards-Anderson type~\cite{BY}:
\begin{equation}\label{EAorder}
q_{EA}(T)=\langle\ \langle s_{\mathbf{x}} \rangle_{T}^2 \rangle_{J},
\end{equation}
where $\langle ... \rangle_{T}$ is the Gibbs canonical statistics thermal average with the Hamiltonian, Eq.~(\ref{RBI}), at temperature $T$, and $\langle ... \rangle_{J}$ is the quenched disorder average with the distribution 
Eq.~(\ref{distri}). However, for many two-dimensional random bond Ising models with a continuous distribution of the bond interactions, like the Gaussian distribution, $q_{EA}(T)=0$~\cite{MorgensternBinder1,McMillanGaussian,BrayMoore,CheungMcMillanGaussian}. This property is due to the presence of soft modes, representing, for the Ising systems, flipped (with respect to the ground state) clusters of connected spins. Also, some sites may be connected to the outside by couplings that all turn out to be small, or the `effective' field from the outside spins may too turn out to be small. This should be contrasted to the homogeneous Ising model on the square lattice, where any excitation with respect to the ground state carries an energy greater than the gap $\Delta=8J$, and the specific heat follows the Arrhenius activation law: $C(T)\approx (\beta\Delta )^2\exp(-\beta \Delta)$, at low temperatures $T=1/\beta$. The average susceptibility of the Gaussian random bond Ising model with respect to the uniform external magnetic field is: $\langle \chi \rangle =1/T$ per site, whereas the low temperature behaviour of the non-linear third harmonics susceptibility $\langle \chi_3 \rangle$ may be more revealing of the ground state properties~\cite{BY}.

Since the energy reference point is physically irrelevant, it is convenient to shift the bond energy by a constant, thus redefining the Hamiltonian as:
\begin{equation}\label{shift}
H'=-\sum_{\langle \mathbf{x}\mathbf{y} \rangle}\left( J_{\mathbf{x}\mathbf{y}} s_{\mathbf{x}} s_{\mathbf{y}} - T \log \left[\cosh\left( \beta J_{\mathbf{x}\mathbf{y}} \right) \right] \right) .
\end{equation}
The thermodynamic averages of the observables can be derived from the partition function, corresponding to the Hamiltonian~(\ref{RBI}): $Z[J]$. The reduced partition function $Z'[J]$ of the model Eq.~(\ref{shift}) is normalized by the total number of configurations $2^N$:
\begin{equation}\label{Zshift}
Z'[J]=\frac{1}{2^N}\sum_{\{s\}}\exp{\left(-\beta H'\right)},
\end{equation}
where the total number of sites is $N$ and the total number of bonds is $2N$. The disorder-averaged free energy per site of the model~(\ref{RBI}) reads: 
\begin{eqnarray}\label{<F>}
F(T) &&= \frac{1}{N} \langle {\cal F}(T,\{J\}) \rangle_{\{J\}} = \frac{1}{N} \langle {\cal F}'(T,\{J\})\rangle_{\{J\}} - \nonumber \\
&&-T \log{2} - 2 T \langle\log\cosh{(\beta J)}\rangle_J, 
\end{eqnarray}
where ${\cal F}'(T,\{J\})=$ $-T\log Z'[J]$ is the total free energy of the model~(\ref{shift}) and the disorder averaging is given by the Gaussian integral Eq.~(\ref{distri}):
\begin{equation}\label{disover}
\langle {\cal F}'(T,\{J\})\rangle_{\{J\}}= \left(\prod_{i=1}^{2N} \int_{-\infty}^{\infty}\!\!\!\! e^{-J_i^2/2} \,\, \frac{dJ_i}{\sqrt{2\pi}}  \right){\cal F}'(T,\{J\}).
\end{equation}
Evaluating this integral for the entire lattice is impossible at present. However, if we are interested in the thermodynamics of a small part of the lattice only, then there are methods available to find the disorder average.

Suppose we know the averaged free energy of the model restricted onto several small clips from the regular lattice. Is it possible to evaluate the averaged free energy for the entire lattice? The graph expansion method answers this question in assertive way. Previously, the linked-cluster expansion method was used to find the high-temperature series for the model Eqs.~(\ref{shift},~\ref{Reduced})~\cite{kadanoff}. In this method the thermal fluctuations, contributing to the free energy in the given order of the inverse temperature $\beta$, are sorted out by their footprint on the lattice, i.e. by the minimal cluster in which a given fluctuation process may occur. It further turns out that many apparently different clusters give identical high-temperature series contributions, and therefore can be combined into broader classes, represented by a single graph, embeddable in the lattice. For that reason we shall call this expansion a graph expansion. This paper's novel approach is to evaluate the free energy of the graph before summing up the graphs. Namely, for each given graph we aim to account for all possible fluctuations in all orders of $\beta$, i.e. we try to find the accurate free energy for each graph in a wide interval of temperatures $\beta$. In this way we order the thermal fluctuations by widening the footprints rather than by growing the perturbation order. A similar approach was developed recently in~[\onlinecite{Rigol}], where a graph property like ${\cal F}_a$ was found numerically by the Hamiltonian diagonalizing procedure rather than by series expansion.

The average free energy per site of the square lattice is given by the graph expansion~\cite{sikes,Aharony}:
\begin{equation}\label{exp}
F(T)= \sum_{a} g_a \langle \delta {\cal F}_a(T,\{J\}) \rangle_{\{J\}},
\end{equation}
where $a$ numerates the connected simple graphs that are embeddable into the square lattice. $g_a$ is the number of embeddings of the graph $a$ into the square lattice. It is comprised of the degeneracy under the action of the point group $D_4$ of the square lattice on the embedded cluster and the degeneracy under the action of all the possible flexes in the joints of the embedded cluster. The cumulant of the graphs' $a$ contribution to the free energy is defined recursively as~\cite{sikes,Aharony}:
\begin{eqnarray}\label{cumulant}
&&\langle\delta {\cal F}_a(T,\{J\})\rangle_{\{J\}}=\nonumber \\
&&\langle{\cal F}_a(T,\{J\})\rangle_{\{J\}}-\sum_{b\in a} g_{ab}
\langle\delta {\cal F}_b(T,\{J\})\rangle_{\{J\}},
\end{eqnarray}
where $\langle {\cal F}_a(T,\{J\}) \rangle_{\{J\}}$ is the average free energy of the Gaussian random bond Ising model, limited to some embedding of the graph $a$ into the square lattice. If $b$ is a subgraph of graph $a$, then $g_{ab}$ is the number of the embeddings of the graph $b$ into any particular variant of the embedding of the graph $a$ into the square lattice. Both $g_{ab}$ and $\langle {\cal F}_a(T,\{J\})\rangle_{\{J\}}$ are independent of a particular variant of the embedding. 
Note, that Eq.~(\ref{exp}) follows from the above expression when $a$ covers a very large part of the square lattice and, therefore, the identity: $\langle \delta {\cal F}_a(T,\{J\}) \rangle_{\{J\}} =0$, holds asymptotically.

For any graph, that has at least one free end, i.e. a vertex with only one incident edge (denoted hereunder a `dangling' edge), the average cumulant of the free energy, evaluated using Eq.~(\ref{Reduced}), is exactly zero. Eliminating such dangling bonds one by one, we end up with a backbone graph, that consists of vertices of incidence degree two, three and four, all embeddable into the square lattice. The cumulant contributions from the graphs that are one-line reducible also vanish. As well as the cumulant contributions from the graphs that can be disconnected by cutting out a four-fold vertex, the so-called articulate graphs. The average free energy may depend only on the total length of the line, and not on its particular embedding route.

A weight of a graph is defined by its number of edges. We enumerate all backbone graphs without dangling edges up to weight fifteen and calculate their lattice degeneracy constants $g_a$, as well as their subgraph constants $g_{ab}$. As the weight of a graph $a$ increases the degeneracy constant $g_a$ grows exponentially. In addition, the sign of the graph cumulant $\delta {\cal F}_a$ changes from one graph to another graph, resulting in the divergent sum Eq.~(\ref{exp}). One way to proceed is to combine graphs into broader classes with the convergent total cumulant contribution. A classification, in terms of graph weights, do not produce a convergent series. In section V we describe a classification of graphs in terms of their envelopes.

One example of the graph expansion is a representation of the famous Onsager formula for the free energy of the regular Ising model on the square lattice:
$$ \iint_0^{\pi} \frac{d\theta d\vartheta}{2 \pi^2} \log\left[(1+x^2)^2 -2x (1-x^2) \left(\cos\theta +\cos\vartheta\right)  \right]=$$
\begin{eqnarray}
=\log \left[(1+x^4) (1+x^6)^2 \left(\frac{1+2x^4+x^6}{(1+x^4)^2(1+x^6)}\right)^2 \cdot \right. \nonumber\\ \left. \left(1+x^8\right)^7 \!\! \left( \frac{1+x^4+x^6+x^8}{(1+x^4)(1+x^6)(1+x^8)}\right)^{12} \!\!\!\! \left(1+x^{10}\right)^{28}\!\! ...\right]\nonumber \\
\end{eqnarray}
in terms of graph polynomials in variable $x=\tanh(\beta J)$ in the rhs, each representing the high-temperature expansion on a particular graph $a$. Powers $1,2,2,7,12,28$ in the rhs count the first $g_a$ constants whereas polynomials in the denominators represent the high-temperature expansion on subgraphs $b$ with their powers being the corresponding $g_{ab}$ constants.

\section{Vanishing density of one-site states at low energy}

At finite temperatures, the Curie-Weiss mean field theory is appropriate for the Ising model, with the excitations being the spin flips on one site in the mean field of the adjacent spins. However, for the two-dimensional Gaussian random bond Ising model at zero temperature the situation is different. In this section, restricted to zero temperature, we will show that the density of the one-site spin flip states vanishes at zero energy.

Consider a site $\mathbf{x}$ with four adjacent sites on the square lattice numerated by an index $i=1..4$. The four bonds adjacent to the site $\mathbf{x}$ with exchange interactions $J_i$ constitute a star. Let $\sigma_\mathbf{x}[J_i,J_j]=\pm$ and $\sigma_i[J_i,J_j]=\pm$ be the value of the central spin and the values of the four tip spins in the ground state. They depend on the disorder configuration $J_i$ on the star as well as on the $J_j$ of all the remaining bonds on the infinite lattice. The excitation energy of a spin flip on the site $\mathbf{x}$ averaged with respect to the disorder outside the star reads:
\begin{equation}
\langle \epsilon_\mathbf{x} \rangle =2\sum_{i=1}^4 J_i C(J_i),
\end{equation}
where the correlation function reads:
\begin{equation}\label{CorrelF}
C(J_i)=\langle \sigma_\mathbf{x}[J_i,J_j] \sigma_i[J_i,J_j] \rangle_{\{J_j\}}
\end{equation}
For a moment we consider the correlation function as being averaged with respect to all the random bonds except one, $J_i$. In this case, if the given exchange coupling $J_i$ is strong enough, the correlation function saturates: $C(J_i)\to \text{sign}(J_i)$. For $J_i=0$ the ground state configuration $\sigma[J]$ is the same as that in the lattice model with a defect - the bond $i$ being cut out. Gaussian symmetry requires $C(0)=0$. For the Gaussian random bond Ising model, defects that break down the translational invariance introduce locally two-level states with a dangling bond attached to the boundary being a one plain example. Those local excitations that flip $\sigma_\mathbf{x} \sigma_i$ are therefore more numerous on average in the presence of the defect, $J_i=0$, compared to the case of no defect, $J_i$ is arbitrary, or, equivalently: $\rho_{local}(\epsilon)>\rho(\epsilon)$. By definition: 
\begin{equation}\label{CorreL}
C(J)=2\int_{0}^J \rho_{local}(\epsilon) d\epsilon .
\end{equation}
For a finite density of local defect states we find: $C(J)=\beta_{eff} J$, at a weak exchange coupling. The mean-field Ansatz, $C(J_i)=\tanh(J_i\beta_{eff})$, interpolates between the strong and the weak limits, where $\beta_{eff}$ is the mean-field bond susceptibility. Returning to the entire star, the correlation function for the one hand of the star will depend also on the remaining random bonds inside the star $J_k$:
\begin{equation}
C(J_i)=\tanh(J_i\beta_{eff}-\sum_{k\neq i} \beta_{ik}' J_k).
\end{equation}
The diagonal elements $\beta_{eff}$ dominate over the off-diagonal elements $\beta_{ik}'$ and the total matrix has all the eigen values positive. We combine all the matrix elements into one $4\times 4$ matrix $\beta_{ik}$. The number of the one-site spin flip excitations with energy less than $\epsilon$ is:
\begin{equation}\label{Number1}
\mathcal{N}_1(\epsilon)=\int \theta\left(\epsilon-2\sum_{i=1}^4 J_i \tanh(\beta_{ik}J_k) \right) \prod_{i=1}^4 e^{ -J_i^2/2} \frac{dJ_i}{\sqrt{2\pi}} .
\end{equation}
For small $\epsilon$ and small $J_i$ it is given by the volume of a four-dimensional ellipsoid, which is an eigen property of the matrix $\beta_{ik}$. Let the surface of this ellipsoid be $S_4$. The density of the one-site states is a derivative of the number of states, Eq.~(\ref{Number1}): 
\begin{equation}\label{DoS1}
\rho_1(\epsilon)=\frac{S_4}{2(2\pi)^2}\ \epsilon,
\end{equation}
and it vanishes at low energy. Note, that in the one-dimensional Gaussian random bond Ising model the density of one-site spin flip states follows the linear law of Eq.~(\ref{DoS1}), whereas the total density of states including extended excitations has a finite limit $\rho(0)\neq 0$.

Let us assume now the power law for all the densities of states under consideration: $\rho_{local}(\epsilon)\propto \epsilon^{\alpha_l}$, $\rho_{1}(\epsilon)\propto \epsilon^{\alpha_1}$ and $\rho(\epsilon)\propto \epsilon^{\alpha}$. Since the local density of states near a defect is greater than without a defect: $\alpha_l\leq \alpha$, and also: $\alpha_1\geq \alpha$, since the one-site excitations are a part of all the excitations. The exponent of the rewritten Eqs.~(\ref{Number1},~\ref{DoS1}) now reads: $\alpha_1=-1+4/(2+\alpha_l)$. If all the three exponents for the density of states are equal, we find:
\begin{equation}
\alpha=\frac{\sqrt{17}-3}{2}.
\end{equation}
The exponent $\alpha_1\geq (\sqrt{17}-3)/2\approx 0.56$ of the one-site density of spin flip states is positive and $\rho_1(0)=0$.

To conclude this section, we find that in two dimensions the one-site excitations are subleading to the extended excitations.

\section{High-temperature series}

Since there is no Edwards-Anderson order at any temperature, the properties of the Gaussian random bond Ising model on the square lattice~(\ref{RBI}) at low temperatures can be approached starting from the well-defined maximum-entropy state at high temperatures. Indeed, in Appendix~\ref{MultiIntegrals} we demonstrate the equivalence of i) the direct disorder average Eq.~(\ref{disover}) and ii) the high-temperature series, when the model is restricted to small lattice clips. In practice, however, the analysis of continuation of the high-temperature series is more efficient.

Let us recall how the high-temperature series is constructed. We transform the partition sum of the model Eq.~(\ref{Zshift}) as follows:
\begin{eqnarray}\label{Reduced}
Z'[J]&=&\frac{1}{2^N}\sum_{\{s\}}\exp{\left(-\beta H'\right)}\nonumber \\
&=&\frac{1}{2^N}\sum_{\{s\}} \prod_{\langle \mathbf{x}\mathbf{y} \rangle} \left[1+s_{\mathbf{x}} s_{\mathbf{y}} \tanh{(\beta J_{\mathbf{x}\mathbf{y}})}\right].
\end{eqnarray}
Here, we use the identity $\exp{(\beta J_{\mathbf{x}\mathbf{y}} s_{\mathbf{x}} s_{\mathbf{y}})}= \cosh{(\beta J_{\mathbf{x}\mathbf{y}})} + s_{\mathbf{x}} s_{\mathbf{y}} \sinh{(\beta J_{\mathbf{x}\mathbf{y}})} $, following from the properties of the Ising spins: $(s_{\mathbf{x}} s_{\mathbf{y}})^2=1$, and from $\cosh$ and $\sinh$ being an even and an odd function, respectively.

Expanding the product over the bonds in the reduced partition function~(\ref{Reduced}), before taking the sum over all the configurations, we find a polynomial of the variables: $s_{\mathbf{x}}$ and $\tanh{(\beta J_{\mathbf{x}\mathbf{y}})}$. A graph on the square lattice can be assigned to each term of this polynomial. It consists of edges (bonds) which end in the neighbouring pair of vertices (sites) connected by this bond. To each edge $\langle \mathbf{x}\mathbf{y} \rangle$ corresponds a factor $\tanh{(\beta J_{\mathbf{x}\mathbf{y}})}$, and to each vertex $\mathbf{x}$ a factor $s_{\mathbf{x}}^{n_\mathbf{x}}$, where $n_\mathbf{x}$ is the incidence degree of the vertex $\mathbf{x}$, i.e. the number of edges which are incident on it. As each lattice site $\mathbf{x}$ has 4 neighbours, so $1\leq n_{\mathbf{x}}\leq 4$. After summing up over all the configurations $s_{\mathbf{x}}=\pm 1$, the terms which contain odd powers of $s_{\mathbf{x}}$ will vanish. As $s_{\mathbf{x}}^0= s_{\mathbf{x}}^2= s_{\mathbf{x}}^4=1$, so each term of the polynomial, containing all the variables $s_{\mathbf{x}}$ in even powers, will give a contribution to the partition function, proportional to the total number of the configurations $2^N$. This fact signifies geometrically that an even number of edges meet at each vertex of the graph. Then, the sum proceeds over closed paths in the graph (denoted `loops' hereunder) only, including loops with self-crossings. And we deduce a formula for the partition function of the Ising model:
\begin{eqnarray}\label{Loops}
Z'[J]=1+\sum_k L_k[J],
\end{eqnarray}
where the loop contribution to the free energy is a product over the loop edges:
\begin{eqnarray}
L_k[J]=\prod_{\langle \mathbf{x}\mathbf{y}\rangle=1}^{m_k}\tanh{(\beta J_{\mathbf{x}\mathbf{y}})},
\end{eqnarray}
with $m_k$ being the number of the edges in the $k$th loop. The smallest loop is the square with four edges.

An important advantage of the high-temperature series is that the disorder average Eq.~(\ref{disover}) is reduced to a straightforward combinatorial using the coefficients: $\langle J_{\mathbf{x}\mathbf{y}}^{2n}\rangle_{J} =(2n-1)!!$. The graph expansion method provides a practical way to produce the high-temperature series for the infinite lattice. We find the high-temperature series of the disorder-averaged free energy for all the graphs up to weight fifteen. And using Eq.~(\ref{exp}), the average free energy per lattice site can be expanded as an exact high-temperature series:
\begin{eqnarray}\label{series}
F(\beta)\! &=& -\log(2)T-\beta +\frac{1}{2}\beta^3 -\frac{2}{3}\beta^5 +\frac{23}{12}\beta^7- \frac{122}{15}\beta^9 \nonumber\\ 
&+& \frac{1786}{45}\beta^{11} -\frac{66364}{315}\beta^{13} +\frac{3085051}{2520}\beta^{15}-  \nonumber\\ 
&-&\frac{22444382}{2835}\beta^{17}+ \frac{813234346}{14175}\beta^{19}-  \nonumber\\ 
&-&\frac{72006710824}{155925}\beta^{21}+ \frac{1898949509689}{467775}\beta^{23}- \nonumber\\ 
&-&\frac{233827938123784}{6081075}\beta^{25}+ \frac{16626101378460212}{42567525}\beta^{27}   \nonumber\\ &-& \frac{2719397636783542268}{638512875}\beta^{29} +   O(\beta^{31}),
\end{eqnarray}
where the energy unit $J=1$ can be reinstated by dimension counting. This series is impractical to analyse at low temperatures as it is too short and irregular. At intermediate temperatures $\beta J\sim 1$ we will use these series as a reference goal when finding the extrapolation from smaller graphs to larger graphs. 

At low temperatures, we evaluate the free energies for each graph that have weight less or equal than fourteen individually and more precisely, namely up to the 127th order of $\beta$. The asymptotic behaviour of this longer high-temperature series justify the Borel transformation, then the series is re-summed using the Pad\'{e} approximation method and the free energy is found by a subsequent Borel integral, see Appendix~\ref{MultiIntegrals} for the explicit procedure. In the temperature interval $0<\beta J< 3$, the thus calculated average free energy for any particular set of graphs turns out to be of high accuracy. Occasionally, but not for the specific sets of graphs used below, the Pad\'{e}-approximants develop pole-like singularities which can be dealt with by finding some longer series. In order to reach even lower temperatures one needs a longer series.

\section{Specific heat and density of states}

Important practical problem besetting the graph expansion method is how to arrange graphs in a meaningful way. We notice, that rearranging graphs by weight up to $W$ results in a partial sum of graphs: $F_W(\beta)$. The high-temperature series of $F_W(\beta)$ in powers of $\beta$ coincides with that of Eq.~(\ref{series}) up to the power $2W-1$, and deviates in higher orders. Unfortunately, we also observe that at low temperatures the functions $F_W(\beta)$ start to increasingly overshoot at $\beta>2$ in opposite directions, for odd and even weights $W=10,11,12,13,14$. There exist mathematical methods for the re-summation of divergent series like the Euler method but after trial and error, we find a physically meaningful way for a proper ordering of the graphs. Namely, we draw a particular embedding of a self-avoiding polygon $l$ on the square lattice and, then, draw progressively more and more lines of edges inside this polygon until all the edges inside will be filled. This procedure gives us a set of graph's embeddings $\{a \}_l$ with the partial sum $\delta F_l(\beta)$ becoming smaller and smaller as the polygon $l$ grows longer and longer. Physically, we count all possible places of excitations and interactions inside a ``lattice clip'' with a boundary $l$. To find $\delta F_l(\beta)$, practically, the lattice constants are split: $g_a=g_{l}+g_{m}+...$ for $l, m, ...\in a$ into different lattice clips $l$, $m$. Arranging the graphs further by the lengths $l$ of the enveloping polygons gives us apparently convergent partial sums of the graph expansion, Eq.~(\ref{exp}).

\begin{figure} \includegraphics[width=0.5\textwidth]{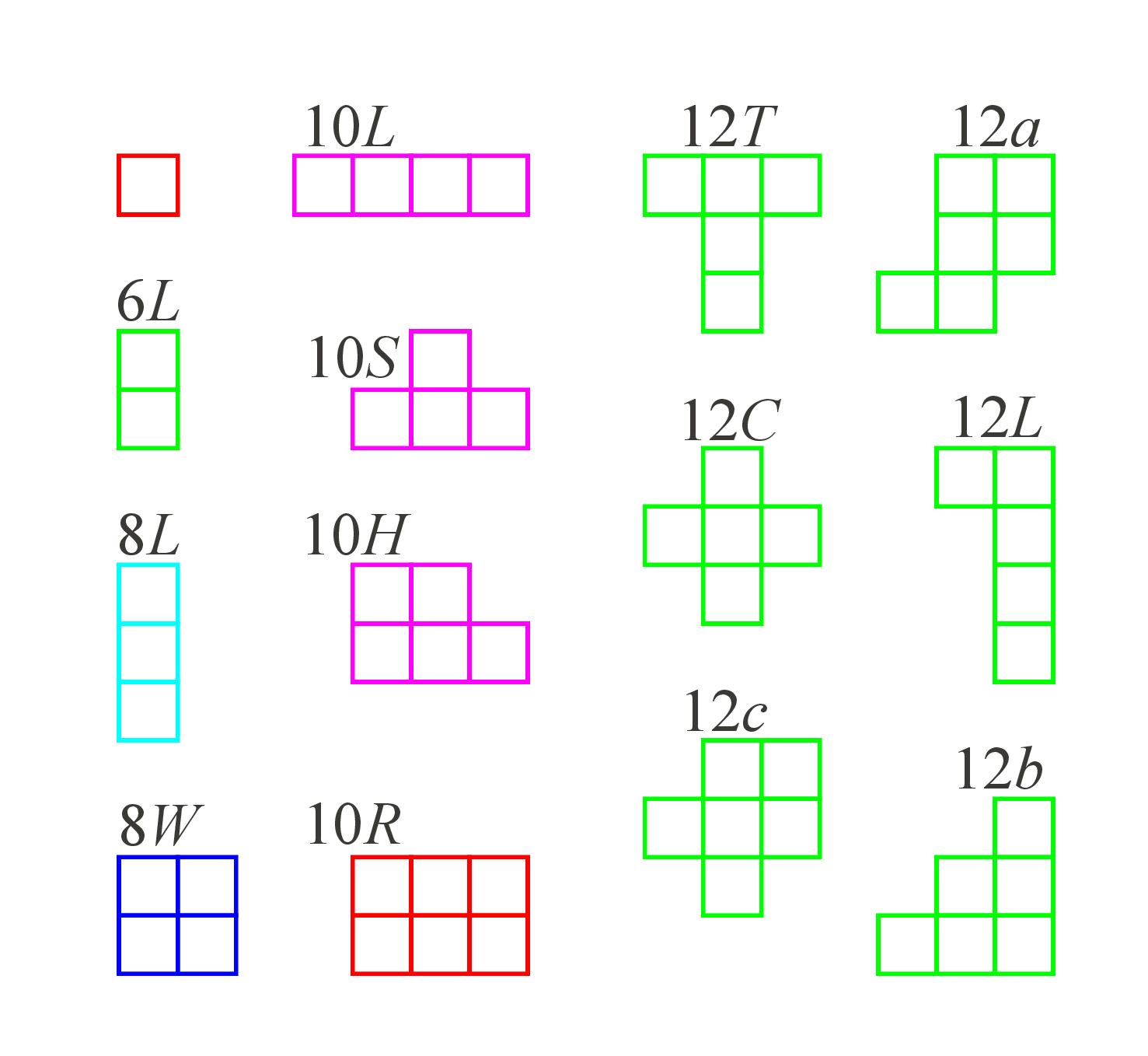}
\caption{Lattice clips used in the graph expansion re-summation, sorted by length of envelope: 4(red), 6(green), 8(blue), 10(magenta and red), 12(green) and shape. Here $L$, $W$, $S$, $H$, $R$, $C$ stand for ``ladder, window, sombrero, hedgehog, rectangle, cross''-like, whereas lower-case a,b,c numerate anonymous graphs of envelope length 12. (Color online.)} \label{diagram}
\end{figure}

\begin{figure}
\includegraphics[width=0.5\textwidth]{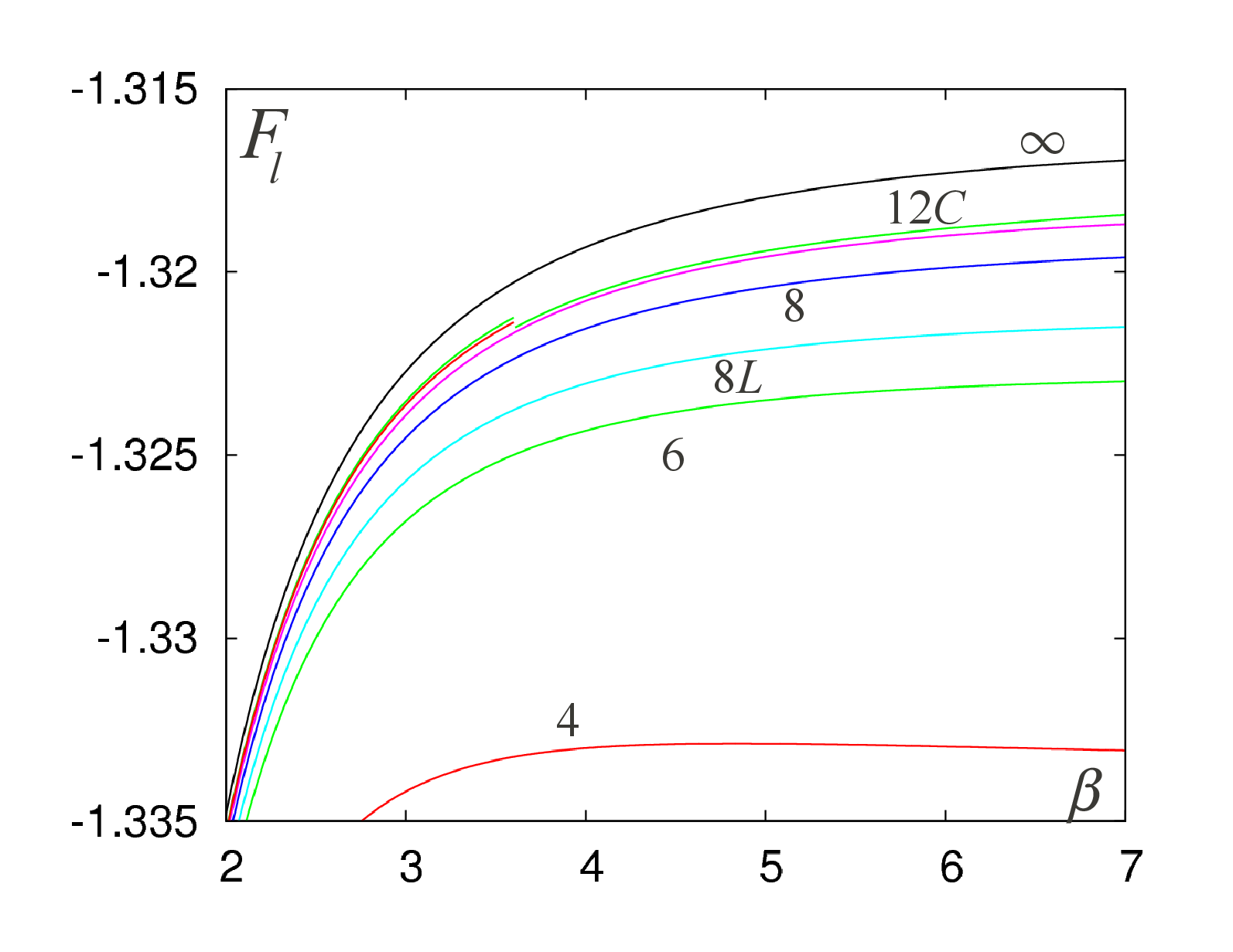}
\caption{The free energy $F_l(\beta)$, in units of $J$, vs the inverse temperature, $J\beta$, for the seven sets of lattice clips with $l=4,6,8L,8,10H,12C,10$, shown in the order from bottom up and in the same colours as in Fig.~\ref{diagram}. The extrapolation of the free energy onto infinite lattice, $F_\infty(\beta)$, is the up-most curve, shown in black. (Color online.)} \label{details}
\end{figure}

We evaluate the average free energy, $F_{l}(\beta)$, for seven sets of graphs with lengths of envelopes no longer than $l=4,6,8,8,10,10,12$. A ``window'' graph, Fig.~\ref{diagram}, of weight $W=12$, is the most ``stuffed'' lattice clip that can be encircled by a polygon of length $l=8$. Accordingly, we split the free energy, corresponding to $l=8$, into two parts: $F_{8L}(\beta)=F_{6}(\beta)+\delta F_{8 L}(\beta)$, belonging to the ``ladder''-like, and $F_{8}(\beta)=F_{8L}(\beta)+ \delta F_{8 W}(\beta)$, the ``window''-like lattice clips, Fig.~\ref{diagram}. Here and below we use a convention that $\delta F_i$ shows the incremental contribution of a given set of graphs, whereas $F_i$ shows the contribution of all the sets of graphs less or equal to the given set. Lattice clips drawn in Fig.~\ref{diagram} illustrate these and the ``sombrero''-like and the ``hedgehog''-like shapes. We can not calculate the average free energy for all the graphs of envelope circumference $l=12$ as there are many difficult ones of weights $W\geq 17$. An especially hard one is the "rectangle"-like lattice clip of envelope length $l=10$, Fig.~\ref{diagram}. We single it out: $F_{10}(\beta)=F_{10H}(\beta)+ \delta F_{10 R}(\beta)$, where the partial sum, $F_{10H}(\beta)=F_{8}(\beta)+ \delta F_{10 L}(\beta)+ \delta F_{10 S}(\beta)+ \delta F_{10 H}(\beta)$, combines the ``ladder''-like, the ``sombrero''-like and the ``hedgehog''-like lattice clips of circumference $l=10$. The next partial sum, $F_{12C}=F_{10}+ \delta F_{12 L}+ \delta F_{12 T}+ \delta F_{12 a}+ \delta F_{12 b}+ \delta F_{12 c}+ \delta F_{12 C}$, includes six lattice clips of envelope length $l=12$, with the last one being the "cross"-like clip. However, since it also includes the "rectangle"-like clip, we can only calculate $F_{12C}(\beta)$ at high temperatures. At low temperatures we exclude the "rectangle"-like clip and use a definition: $F_{12C}=F_{10H}+ \delta F_{12 L}+ \delta F_{12 T}+ \delta F_{12 a}+ \delta F_{12 b}+ \delta F_{12 c}+ \delta F_{12 C}$. For each of the seven sets of graphs: $l=4,6,8L,8,10H,10,12C$, the high-temperature series up to the 721t order for $l=4,6,8L$, up to the 481t order for $l=8,10H,12C$ and up to the 177th order for $l=10$ are found. Then, the Borel transformation is applied to these series, followed by a Pad\'{e} approximation and, in the end, the free energy is found using the Borel integral, see Appendix~\ref{MultiIntegrals} for details. Thus found free energies turn out to be already converged enough to be almost independent on the increasing of the maximum order further in the interval $0<\beta J<7$, except for $F_{10}(\beta)$, which is accurate only if $0<\beta J<3.6$. The free energy curves as functions of the parameter $l$ are shown in details in Fig.~\ref{details}, in the same colour as they appear in Fig.~\ref{diagram}, and are clearly converging. For the free energy, $T=J/7$ is the lowest temperature we can accurately access.

\begin{figure}
\includegraphics[width=0.24\textwidth]{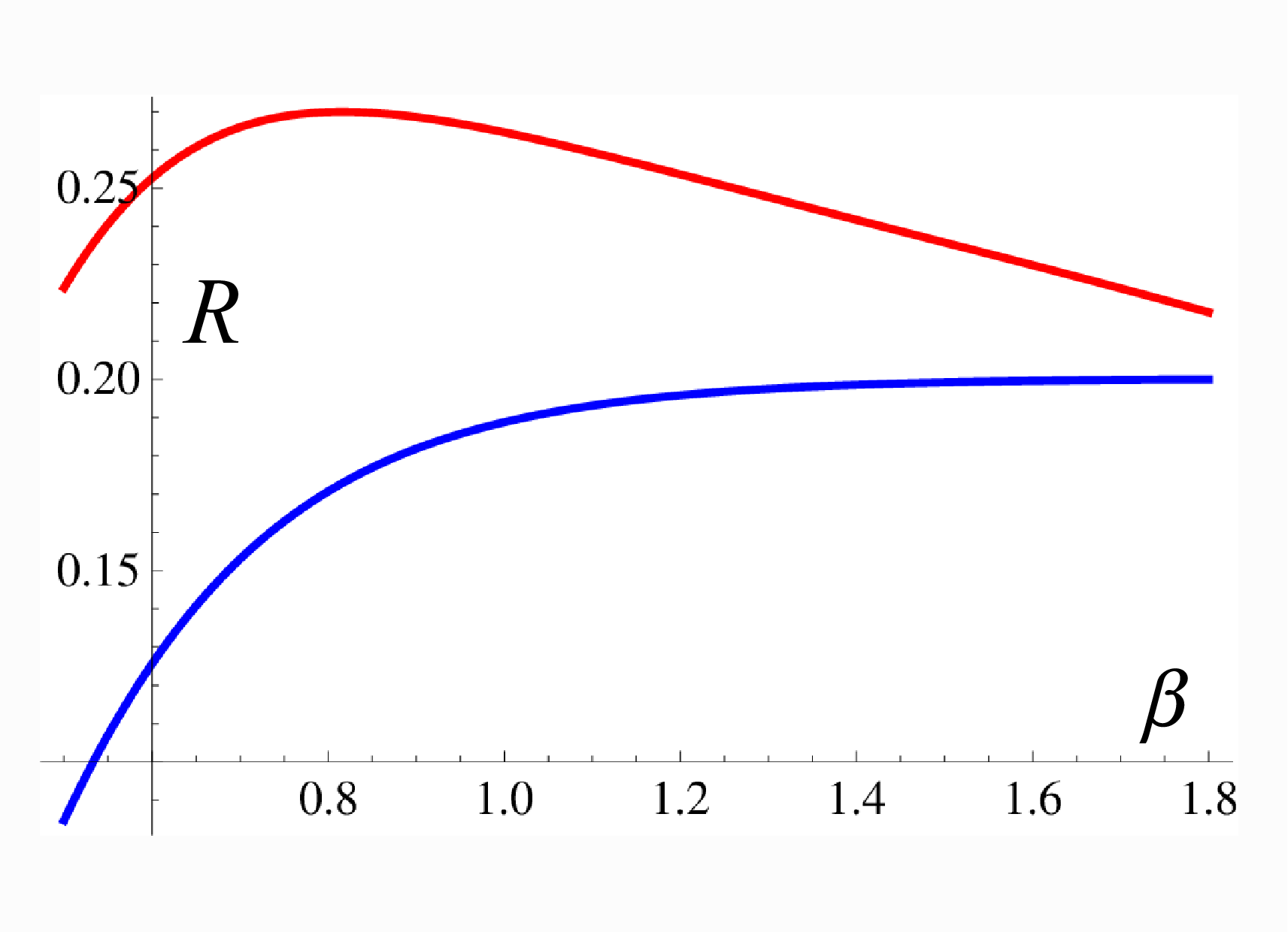}\qquad\qquad\qquad\qquad
\caption{The ratio $R(\beta)$, Eq.~(\ref{R}), the lower blue line and the ratio $(F_{12C}-F_8)/(F_{\infty}-F_{8L})$, the upper red line, as functions of the inverse temperature $\beta$. (Color online.)} \label{extrapolation}
\end{figure}

What is the extrapolation function $F_{\infty}(\beta)$ of the free energy when we add all the larger graphs with longer envelopes? The first chunk of additional larger graphs of weights $W=12,13,14,15$ is already calculated in Eq.~(\ref{series}), which is exact up to the order shown. Applying the Borel-Pad\'{e}-approximation to the exact short series, Eq.~(\ref{series}), produces the function $F_\infty(\beta)$, which is only accurate in the interval $0<\beta<1.4$, but diverges from the exact free energy at lower temperatures. At high temperatures all the differences $F_\infty(\beta)-F_l(\beta)$ are small and approach zero according to $\beta^{2l-1}$ as $\beta\to 0$. We examined many ratios of such differences for the sets of graphs arranged by their envelopes: $F_{12C}$, $F_{10H}$, $F_{8}$ and $F_{8L}$. In Fig.~\ref{extrapolation} two of such ratios are shown. One of them, namely
\begin{equation}\label{R}
R(\beta)=\frac{(F_\infty-\frac{5}{4}F_{12C}+\frac{1}{4}F_8)(F_{12C}-F_8)}{(F_\infty-F_8)^2}, 
\end{equation}
apparently saturates to a constant value $R(\beta)\approx 0.2$ at $\beta> 1$. We assume that the relationship $R(\beta)=1/5$ holds at lower temperatures as well. Solving it for the function $F_\infty(\beta)$ produces the extrapolation of the free energy of the Gaussian random bond Ising model:
\begin{equation}\label{extrapol}
F_\infty(\beta)=\frac{5}{2} F_{12C}(\beta)-\frac{3}{2} F_8(\beta),
\end{equation}
shown in Fig.~\ref{details}. The extrapolation~(\ref{extrapol}) commutes approximately with the Borel-Pad\'{e}-approximation, e.g. it makes little difference if the extrapolation~(\ref{extrapol}) is applied to the original high-temperature series followed by the Borel-Pad\'{e}-approximation or if the extrapolation~(\ref{extrapol}) is applied to the results, functions, of the Borel-Pad\'{e}-approximation. Extrapolation~(\ref{extrapol}) sacrifices the accuracy at temperatures, already high and where $R\neq 1/5$, to improve the accuracy at low temperatures.

Our best estimate of the average free energy per site of the Gaussian random bond Ising model on the square lattice $F_\infty(\beta)$, Eq.~(\ref{extrapol}), is fitted in the interval $2.3<\beta J <6.9$, using the following ad-hoc formula: 
\begin{equation}\label{freeE}
F(T)=-\left((AT)^{2+\alpha}+|E|^{2+\alpha}\right)^{1/(2+\alpha)}.
\end{equation}
This expression is a simple one and is the best to represent our free energy data. It extrapolates between the ground state energy, $F(0)=E$, and the critical scaling law for the specific heat, $C\propto T^{1+\alpha}$, at zero temperature, and the high-temperature behaviour $F=-\log(2) T$, corresponding to a bunch of independent spins. However, it misrepresents the high-temperature specific heat $C\propto T^{-2-\alpha}$, instead of the correct behaviour $C\propto 1/T^2$, and may miss next to the leading crossover term in the description of the zero-temperature criticality. We find, by best fitting 461 points of $F_{\infty}(\beta)$ with the overall accuracy of 0.0006\%, that $A=0.7356$, which is indeed close to $\log(2)$, $\alpha=0.591$ and the average ground state energy per site $E=-1.3162J$. The negative ground state energy is somewhat higher in comparison with $E=-2J$ of the homogeneous Ising model. This effect is due to the frustration and can be understood as follows. All the bonds on the square lattice can be divided into two types, relaxed and frustrated ones, according to the value of $\textrm{sign} (J_{\langle \mathbf{x} \mathbf{y} \rangle} s_{\mathbf{x}} s_{\mathbf{y}} )$ in the ground state configuration of the spins. The frustrated bonds have relatively smaller interactions $|J_{\langle \mathbf{x} \mathbf{y} \rangle}|$, whereas the relaxed bonds have relatively larger interactions $|J_{\langle \mathbf{x} \mathbf{y} \rangle}|$ in the ground state.

\begin{figure}
\includegraphics[width=0.368\textwidth]{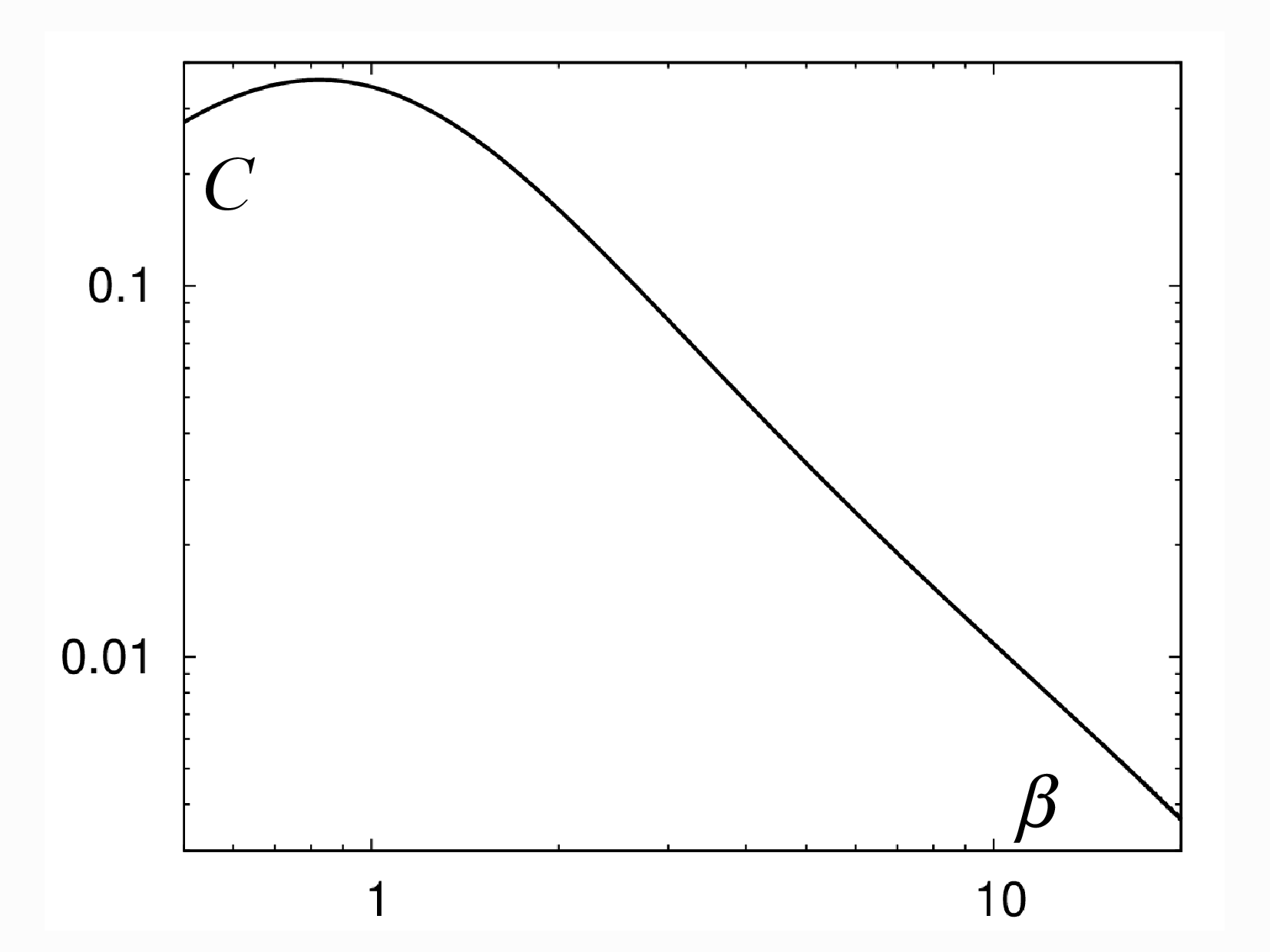}
\qquad\qquad\qquad\qquad
\caption{A log-log plot of the specific heat $C(\beta)$ as a function of the inverse temperature $\beta$. In a wide interval $6<\beta J<20$ a power law sets in.} \label{SH}
\end{figure}

One of the advantages of the graph expansion method is that the high-T series obtained, for, say, the free energy, can be analytically manipulated. We use the formula:
\begin{equation}
C(\beta)=-\beta \frac{d}{d\beta} \beta^2 \frac{dF}{d\beta},
\end{equation}
to produce a high-T series for the specific heat, with the original series for $F$ given by Eq.~(\ref{extrapol}). Borel-Pad\'{e}-approximations of varying nominator and denominator polynomials all give almost the same function $C(\beta)$ in the interval $0<\beta J<20$. The specific heat, therefore, can be approximated to lower temperatures than the free energy. In part this is due to the absence of the $-T\log(2)$ term in the specific heat, as opposed to the free energy. In Fig.~\ref{SH} we show the specific heat as a function of the inverse temperature using the log-log plot. A power law is observed and the best fit gives: 
\begin{equation}\label{SHresult}
C(T)= c\ T^{1+\alpha}, \qquad \alpha=0.55(8),
\end{equation}
where $c \approx 0.40$. Integrating numerically the specific heat function over all temperatures:
\begin{equation}
E=-\int_0^\infty C(T)dT,
\end{equation}
gives the ground state energy per site: $E\approx -1.3160(16)$, agreeing with previous findings~\cite{CampbellHartmannKatzgraber, Thomas_GroundStateEnergy}. The systematic errors to the values of $\alpha$ and $E$ are estimated using the comparison with the Borel-Pad\'{e}-approximation of the specific heat, $C_{12C}(T)$, on the largest non-extrapolated lattice clips, and the comparison with the free energy results, Eq.~(\ref{freeE}). To conclude, we find a zero-temperature critical behaviour for the specific heat, Eq.~(\ref{SHresult}), in the Gaussian random bond Ising model on the square lattice.

One microscopic theory for the low temperature free energy and the specific heat~\cite{AndersonHalperinVarma} is the model of non-interacting two-level states, the droplets. An excited state in an Ising system is specified by a set of connected clusters of flipped spins with respect to the ground state. Mutually disconnected clusters contribute independently to the total excitation energy. The excitation energy of a flipped cluster is positive, and is equal to the energy of the domain wall boundary. The domain wall energy is the sum of the energies of the local bonds, positive and negative alike. With a small probability, we can find a specific fractal-like loop with many negative bonds across it and, if the spins inside are overturned, the domain wall energy will be atypically small, although positive. Such loops are rare in the space of random walks, but in real space are dense. The two-level states are also dense but taken as independent. The free energy of a system of independent two-level states reads~\cite{AndersonHalperinVarma}:
\begin{equation}\label{twoL}
F(T)=E-T\int_0^\infty\rho(\epsilon) \log\left(1+e^{-\epsilon/T}\right)d\epsilon,
\end{equation}
where $\rho(\epsilon)$ is the density of states, and $E$ is the ground state energy. Working out the relationship between the free energy, Eq.~(\ref{freeE}), and the two-level states model, Eq.~(\ref{twoL}), we find the low-energy asymptote for the density of states:
\begin{equation}\label{DoS}
\rho(\epsilon)= \frac{A^{2+\alpha}}{(2+\alpha) |E|^{1+\alpha}} \frac{\epsilon^{\alpha}}{(1-2^{-1-\alpha}) \Gamma\left(1+\alpha\right) \zeta\left(2+\alpha\right)}.
\end{equation}
This result, Eq.~(\ref{DoS}), shows that the excitation spectrum of the Gaussian random bond Ising model on the square lattice differs from i) the same model defined in one dimension and ii) other systems, that develop the spin glass phase at finite temperatures.

\section{Discussion}

\begin{figure}
\includegraphics[width=0.3\textwidth]{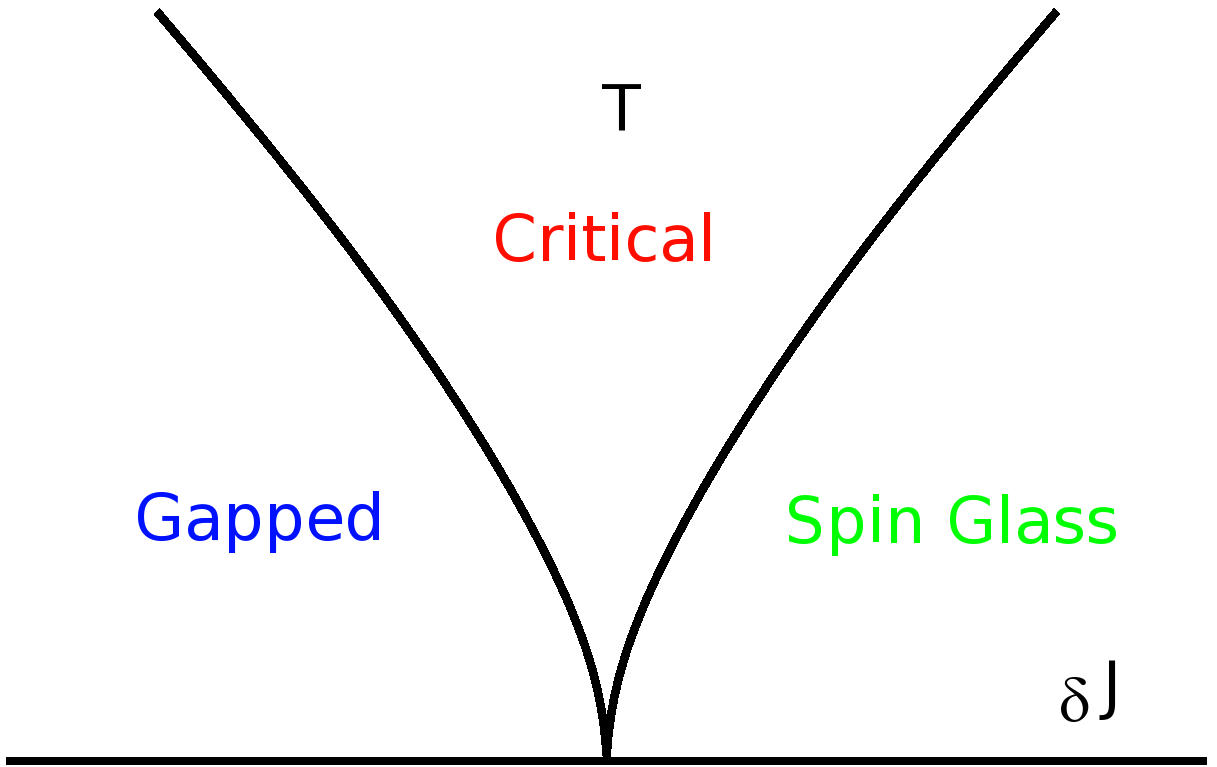}
\caption{A possible phase diagram of the Gaussian random bond Ising model criticality. The vertical axis is the temperature. The horizontal axis is a special variation of the exchange couplings as explained in the text. (Color online.)} \label{criticality}
\end{figure}

The average free energy and the specific heat of the Gaussian random bond Ising model on the square lattice have been found with high accuracy in a wide range of temperatures. At low temperatures both show power law criticality, with $C(T)\propto T^{1+\alpha}$. A convergent graph expansion ordering was found for this model, which may aid in improving the results when using more efficient algorithms for the most difficult graphs.

Graphs used in this study, Fig.~\ref{diagram}, are tiny when compared to clusters of size $L=128$ or so, used in the Monte Carlo methods. Our method evaluates observables directly in the thermodynamic limit of the infinite lattice and the sizes of the graphs are indicative of the correlation length accessible. Unlike it, in the Monte Carlo simulations the boundary effects, the systematic distortions, have to decay below the required accuracy on the lattice size, thus, limiting the accessible correlation lengths.

Our specific heat $C(T)= cT^{1+\alpha}$ deviates from the specific heat $C(T)= aT$ found in~[\onlinecite{HoudayerHartmann}] using the Monte Carlo method. Here we give an account for this discrepancy. A plausible phase diagram of zero-temperature criticality may look like in Fig.~\ref{criticality}. Suppose we find an exact ground state of some large cluster of the Gaussian random bond Ising model. There will be a critical density of the low energy states, Eq.~(\ref{DoS}). By varying all the exchange couplings slightly but adjusting the amplitude and the sign, we can push all these low energy states to even lower energies. It is possible that in this way we can enhance the density of excited states to a small finite value at the zero energy: $\rho(0)\neq 0$. This signals a transformation of the cluster state into a spin glass. If we now change the sign of this unique variation: $\delta J[J]$, which is a functional of a given distribution of $J$, we push the excitations to a higher energy and will 
deplete the density of the low energy states, thus, creating some gapped state. We normalize the variation $\delta J[J]$ by a scalar measure $h=||\delta J[J]||$. Equating the specific heats of the two phases on the spin glass transition line we find $a=cT^\alpha_{SG}(h)$. For the zero-temperature criticality a scaling law $T_{SG}\propto h^\nu$ holds, with the exponent $\nu$ being less than one, and typically $\nu\approx 0.5$. Thus, $a\propto h^{\alpha\nu}$. In the Monte Carlo method a particular realization of the exchange disorder $J$ may fluctuate away from the Gaussian one by a large-number value $h\propto 1/\sqrt{N}=1/L$. It may be difficult in practice to tell apart $a\propto 1/L^{\alpha\nu}$ from $a=const$ if $\alpha\nu$ is small enough, say $\alpha\nu=0.3$. Also, our zero-temperature criticality hypothesis explains why the observed $a$ is so small \cite{HoudayerHartmann}.

The low-temperature specific heat of the Gaussian random bond Ising model on the square lattice was predicted in~[\onlinecite{JLMM}] to be dominated by the regular term $C(T)\propto T$, with the singular, hyper-scaling term $\propto T^{-2/\theta}$ being a sub-leading correction. Recently in~[\onlinecite{A}] a hypothesis that the two-dimensional Gaussian random bond Ising model can be described by a conformal theory, as well as the exact value of $\theta$, was proposed. The conformal theory is scale free. We note, that a small remnant density of states $\rho(0)$ advocated in~[\onlinecite{JLMM}] presents a well defined scale $1/\sqrt{\rho(0) T}$, i.e. the average spatial distance between the two-level excitations. Any such divergent scale, smaller than the correlation length $\xi=T^{1/\theta}$, would contradict the conformal invariance proposed in~[\onlinecite{A}].

\textbf{Acknowledgment.}
We are thankful to Bahcesehir University for their hospitality, where part of this work was completed. We are indebted to Creighton Thomas and Alan Middleton, who communicated us their unpublished data on the same Gaussian random bond Ising model on the square lattice using the Pfaffian method; and to Anastasios Malakis for illuminating discussions.

\appendix

\section{}
\label{MultiIntegrals}

\begin{figure}
\includegraphics[width=0.35\textwidth]{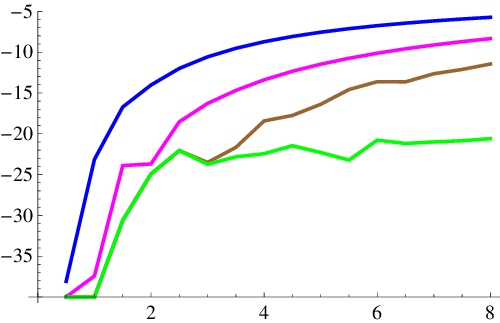}
\caption{The error $R_i(\beta)$ as a function of the inverse temperature $\beta$ for the high-temperature series of lengths: $i=40$, the upper blue line, $i=80$, the magenta line, $i=160$, the brown line and $i=320$, the lower green line. (Color online.)} \label{errors}
\end{figure}

The high-temperature series for the Gaussian random bond Ising model on the square lattice:
\begin{equation}\label{Series}
F(\beta)=T\sum_{n=0}^\infty a_n \beta^{2n},
\end{equation}
shows the asymptotic behaviour of the coefficients: $a_n\to (-1)^n\ (n+5)!$, as $n\to \infty$, and bears a close similarity with the celebrated Borel re-summation example:
\begin{equation}
\sum_{n=0}^\infty (-1)^n\ n! \beta^{2n}=\int_0^\infty \frac{1}{1+z\beta^2}e^{\displaystyle -z}\ dz.
\end{equation}
Nevertheless, the series Eq.~(\ref{Series}) has a zero radius of convergence and a question arises whether the re-summation of the high-temperature series faithfully represents the averaged free energy. In this Appendix we will demonstrate the equivalence of the two ways to evaluate the average free energy on a finite graph, specifically in the simplest case of a one square graph with four edges and four nodes. The first approach uses the direct calculation of the disorder integral:
\begin{eqnarray}\label{multi}
\langle F(\beta)\rangle =-T \int \frac{dJ_1 dJ_2 dJ_3 dJ_4}{(2\pi)^2}\,
e^{-(J_1^2+J_2^2+J_3^2+J_4^2)/2}\nonumber \\
\log \sum_{s_1 s_2 s_3 s_4=\pm} e^{\beta (J_1 s_1 s_2 + J_2 s_2 s_3 +J_3 s_3 s_4 +J_4 s_4 s_1) }.\nonumber \\
\end{eqnarray} 
The second approach employs the expansion of the disorder-averaged free energy into the high-temperature series of $i$ terms followed by the Borel transformation and the Pad\'{e} approximation:
\begin{equation}
F(\beta)\to \sum_{n=0}^i \frac{a_n}{(n+k)!} \beta^{2n} \to \frac{P_i(\beta)}{Q_i(\beta)},
\end{equation}
where $P_i(\beta)$ and $Q_i(\beta)$ are polynomials of the order of or around $i$, and the number $k$ can be varied. Finally, the Borel integral renders the free energy:
\begin{equation}
F(\beta)\approx F_i(\beta)=\int_0^\infty dz\ z^k\frac{P_i(\beta\sqrt{z})}{Q_i(\beta\sqrt{z})} e^{\displaystyle -z}.
\end{equation}
In this way we evaluate four approximations for the free energy: $F_{i}(\beta)$, for series of $i=40,\ 80,\ 160,\ 320$ terms. These four are compared with the multi-dimensional integral value Eq.~(\ref{multi}) in terms of the error function:
\begin{equation}
R_i(\beta)=\log\left|\frac{F_{i}(\beta)-\langle F(\beta)\rangle}{\langle F(\beta)\rangle}\right|.
\end{equation}
We plot the error $R_i(\beta)$ in Fig.~\ref{errors} as a function of $\beta$ for $i=40,80,160,320$ terms ordered from up to bottom. For medium temperatures $\beta\sim 1$ both methods are accurate up to $\exp(-30)$ or better. Here, the accuracy of the high-temperature series is much better than the accuracy of the multi-dimensional integral Eq.~(\ref{multi}), resulting in a somewhat erratic picture. For low temperatures $\beta\gg 1$, the accuracy of the high-temperature series deteriorates faster than the multi-dimensional integral Eq.~(\ref{multi}). However, by increasing systematically the length of the high-temperature series one can improve the accuracy up to $\exp(-20)$. 

In conclusion, there is no systematic obstruction for the application of the high-temperature series, and, moreover, proceeding with the high-temperature series turns out to be a more efficient way than taking the multi-dimensional disorder integrals.

\end{document}